\documentclass[a4paper]{jpconf}
\usepackage{graphicx}
\begin{document}
\title{  
  Deuteron production in p-Be interactions at 450 GeV/c
  and the coalescing model}
\begin{center}
\vskip 0.5cm
{\large NA56 Collaboration}
\end{center}
\author{M. Bonesini$^1$}

\address{$^1$ Sezione INFN Milano Bicocca, Dipartimento di Fisica 
G. Occhialini, \\
Piazza Scienza 3, Milano, Italy}

\ead{maurizio.bonesini@mib.infn.it}

\begin{abstract}
The analysis of the deuteron production  in p--Be interactions at 450 
GeV/c taken by the NA56/SPY experiment at CERN SPS is presented.
In the framework of the coalescence model,  
the coalescence factor $\kappa$ is determined as $(0.79 \pm 0.05 \pm 0.13) \times
10^{-2}$. 
Our results disfavour the hypothesis that coalescence 
be the dominant mechanism for deuteron production in $p+Be$ interactions 
at low $p_T$.
\end{abstract}
At high collisions energies 
($E_p \sim \ 1 \ GeV$), the cascading of the incident proton and the 
secondary particles inside the nuclear target can produce several 
fast nucleons. In a simple {\it coalescence model} \cite{Butler}
these independent nucleons may fuse to a deuteron due to final-state
interactions,
if their relative momenta are within 
a momentum sphere of radius $q_0$. 

In the limit $p_d \leq p_{d}^{lim}=0.4 \times p_{inc}$,
the deuteron production cross section is given by:

$$ E_{d}\frac{d^{3}\sigma}{d p_{d}^3} =
\frac{\kappa \times R}{\sigma_{in}}(E_{p}\frac{d^{3}\sigma}{d p_{p}^3}) \times
(E_{n}\frac{d^{3}\sigma}{d p_{n}^3}) \simeq
\frac{\kappa \times R}{\sigma_{in}}(E_{p}\frac{d^{3}\sigma}{d p_{p}^3})^2  \ \ \ \ (1)$$

where $\kappa$ is the coalescence factor, $R$ the two-particle correlation
function and $\sigma_{in}$ the proton-nucleus
cross section\footnote{In the second part of equation (1) it has been
assumed that $p_p=p_n=p_d/2$ and the cross section for proton production has been
put equal to the one for neutron production}.
In \cite{Butler} it was shown that
$q_0=150-200$ MeV/c and that $\kappa \sim q_0^3$.
The quantity $R$ is most frequently taken equal to unity, which corresponds
to statistically independent production of the particles making up a pair.

The NA56/SPY collaboration \cite{SPY} has collected data with 450 GeV/c
protons hitting beryllium targets of different lengths and shapes.
Data have been collected over a secondary particle momentum range from 
7 GeV/c to 135 GeV/c and up to 600 MeV/c transverse momentum. 

The experimental apparatus consisted of the NA52 spectrometer in the 
H6 beamline 
at CERN SPS equipped with proportional chambers for tracking and
TOF stations, Cherenkov detectors and a hadron calorimeter for particle 
identification, as described in reference~\cite{analisi}. 

The $d/p$ production cross-section ratio 
was extracted from the data taken with
a 100 mm long Beryllium target and 
combined with the $p$ cross section,
as determined in \cite{analisi}, was used to obtain the deuteron 
cross section.

In the measurement of $d/p$ production ratio, systematics are due to 
particle dependence of the transmission along the H6 beamline ($\sim 2 \%$),
uncertainties in the subtraction of $\Lambda \mapsto p \pi^-$ decaying 
outside the target (conservatively a $\sim 30 \%$ error was assumed on 
the value of the correction \footnote{The contamination of 
$\Lambda \mapsto p \pi^{-}$ goes from $14.4 \%$ at 7 GeV/c to 
$5 \%$ at 40 GeV/c.}) and 
empty target corrections. 

The systematic uncertainties of deuteron cross-section
 include in addition the uncertainties in the 
knowledge of primary beam intensity ($\sim 1.7 \%$), 
the spectrometer acceptance and particle transmission
along the beamline ($\sim 10 \%$).

Protons are identified, using similar selection criteria as applied 
in \cite{analisi}. Deuterons are distinguished from protons by the
TOF reconstruction up to 40 GeV.
At higher energies the separation $p/d$ is marginal.
Target efficiency for protons and deuterons was computed with a naive absorption
model \cite{malensek}, assuming for the computation of the interaction
lengths the nuclear cross section given in reference \cite{Bamberger}.
Particle losses along the beamline were calculated by means of an updated 
version of the TURTLE Montecarlo simulation of beam transport
\cite{turtle}, with inclusion of multiple scattering and nuclear interactions
in the detector and the beam material. 


Results on the particle cross-section ratio (d/p) in the forward direction, 
as a function  of beam momentum and at fixed momenta (+15, +40 GeV/c) as 
a function of $p_T$ are shown in figure 1.
All numbers are corrected for $\Lambda \mapsto p \pi^-$ decays outside 
the target. Systematic errors have been added in quadrature to the 
statistical ones in the reported plots. 
The systematic uncertainty due to the different transmission of $d$
 and $ p$
along the H6 beamline has been treated as a common systematics in the
production angle scans and not included in the total reported systematic errors.

For comparison the results obtained by A. Bussiere et al. \cite{Bussiere}
in p-Be collisions at 200 GeV/c are plotted.
Due to the difference in primary beam momentum (200 GeV/c instead of
450 GeV/c), their results have been plotted at $(450/200) \times p_{d}$ 
in order to compare the same $x_F$ values.
As already observed in other experiments, deuterons are more copiously
produced with respect to lighter particles at greater $p_T$ values.

\begin{figure}[hbtp!]
\begin{center}
\hspace{-0.18cm}
\mbox{\includegraphics[width=.25\linewidth]{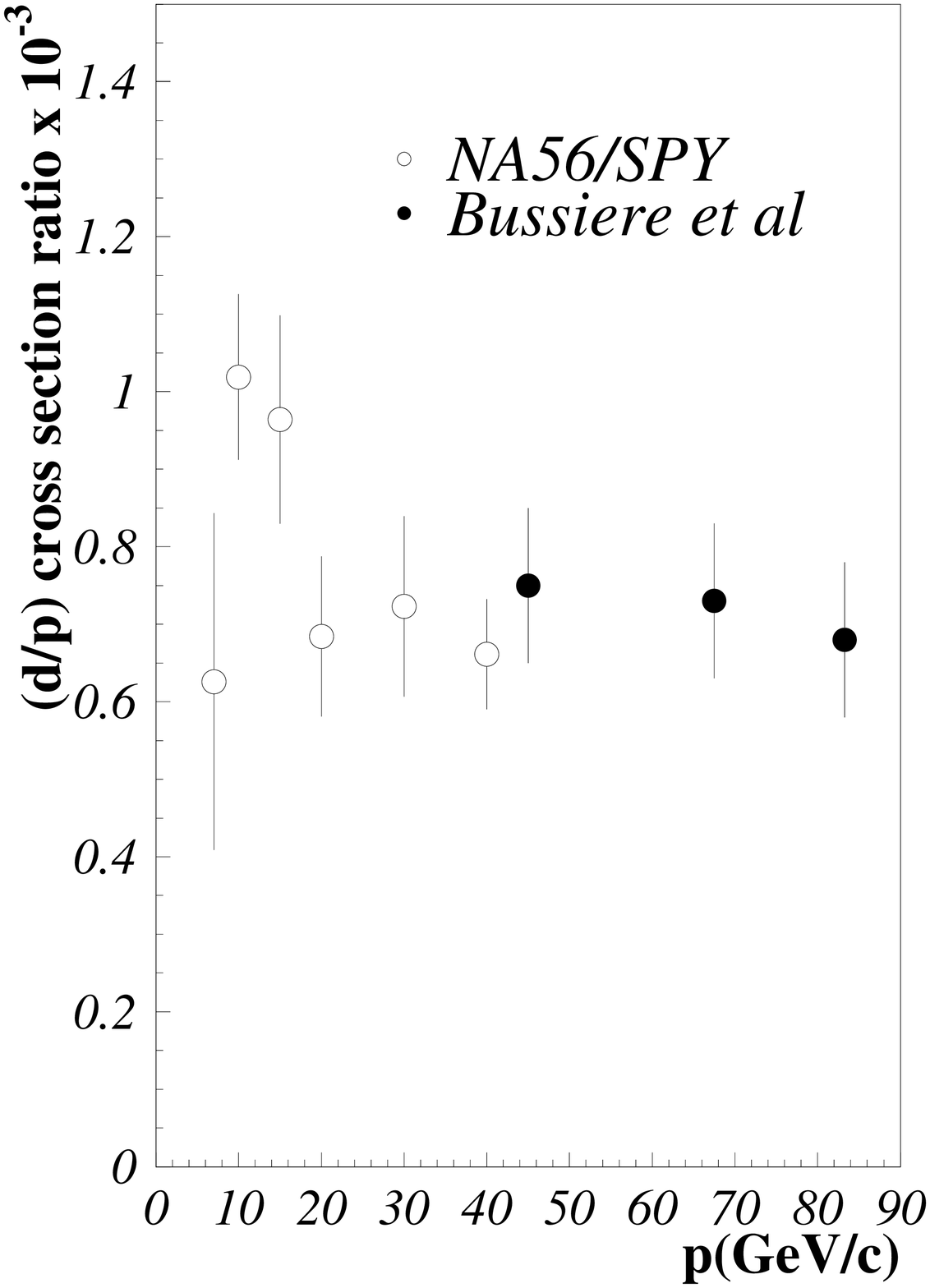}} 
\hspace{-0.4cm}
\mbox{\includegraphics[width=.25\linewidth]{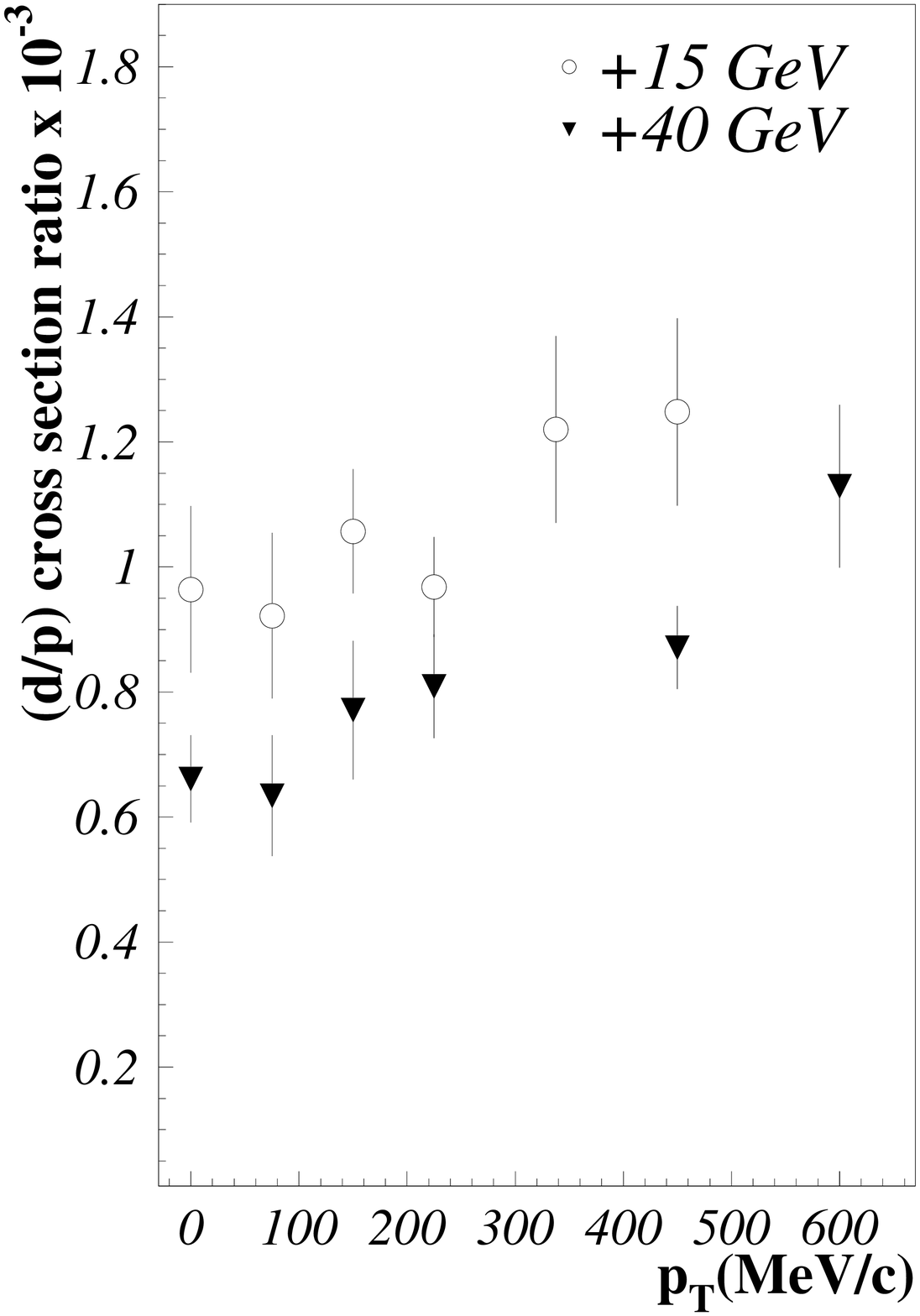}} 
\mbox{\includegraphics[width=.25\linewidth]{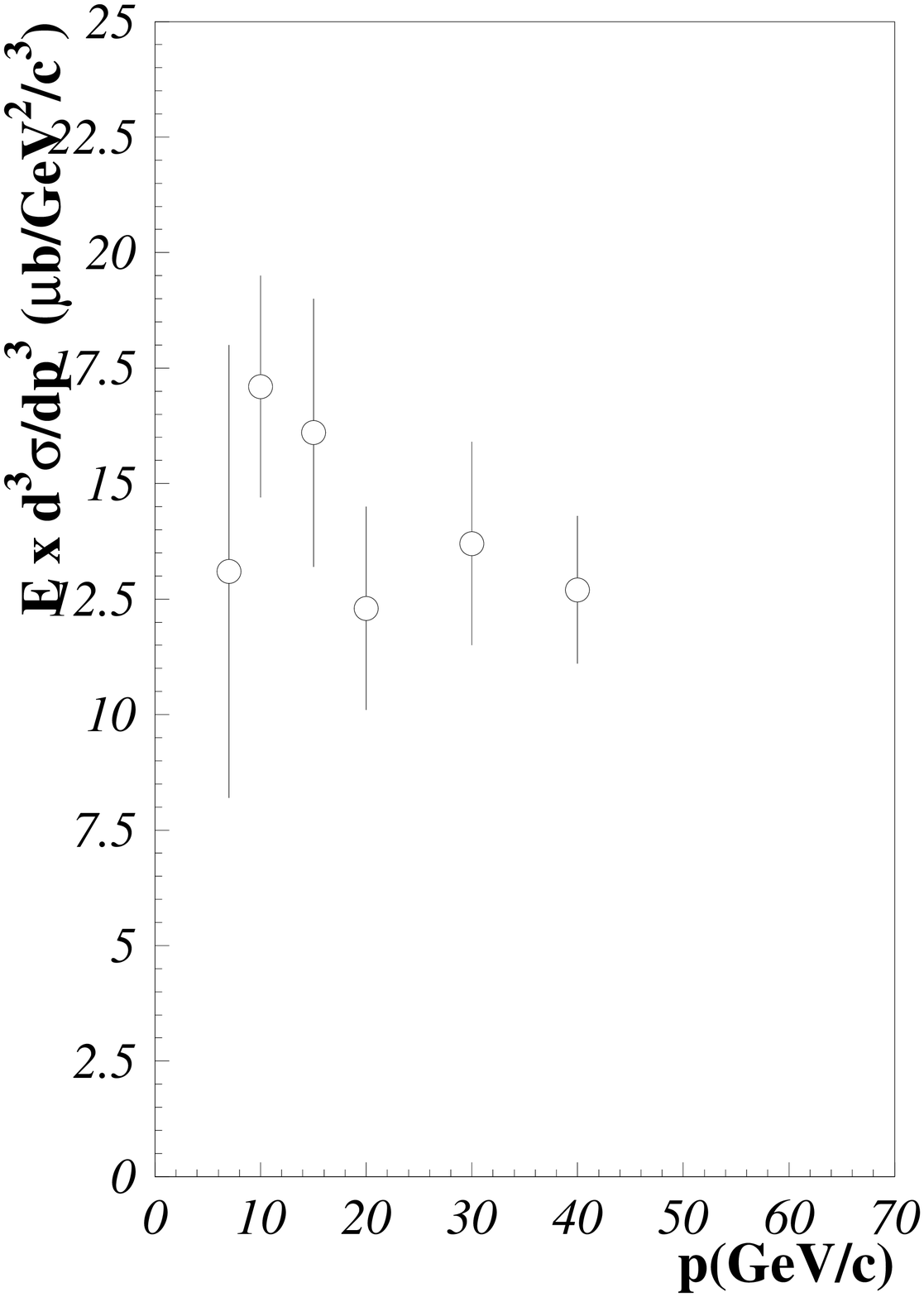}}  
\mbox{\includegraphics[width=.25\linewidth]{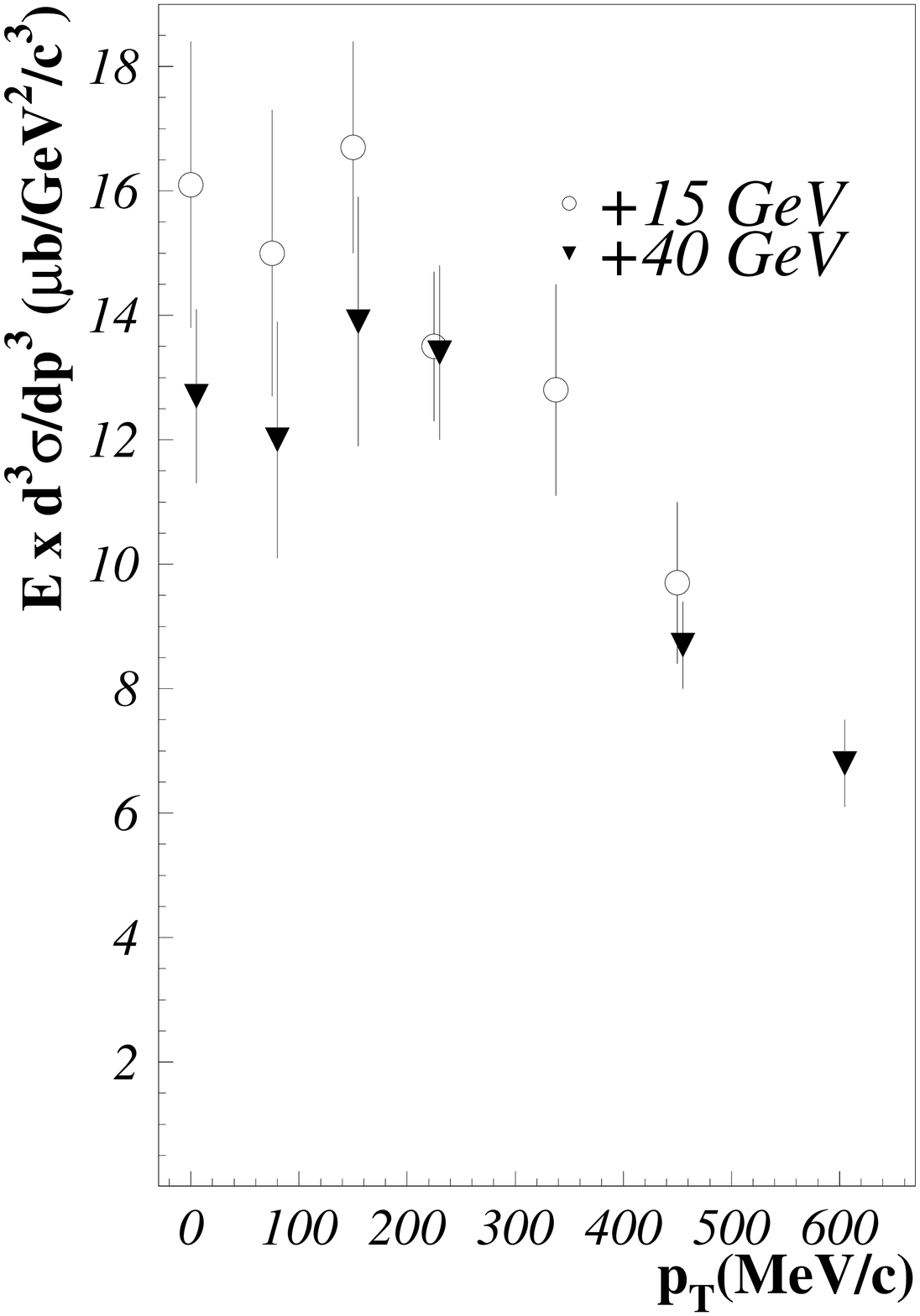}} 
\end{center}
\caption{ From left to right: $d/p$ cross-section production ratio as a function of momenta;
$d/p$ cross-section production ratio as a function of $p_T $ at +15 GeV/c
and at +40 GeV/c;
  $d$ invariant cross section as a function of 
momentum, in the forward direction; d invariant cross section 
as a function of $p_T$
at +15 GeV/c and +40 GeV/c.}   
\label{f16a}
\end{figure}

The deuteron invariant cross section has been obtained combining the 
$(d/p)$ ratio with the proton invariant cross section, as measured in \cite{analisi}. The dominant systematic
error is due to the acceptance calculation, that is considered
a common systematics in the angular scans (and is not included in the
total errors in this case). 
Results are also shown in figure 1.
Systematic errors have been added to statistical errors in the reported plots.

In the  coalescence model for deuteron production, 
assuming a two particle correlation function $R$=1, the
coalescence factor $\kappa$ is given by:
$\sigma_{in} \frac{(E d^{3}\sigma/dp^{3})_d}{(E d^{3}\sigma/dp^{3})_p^{2}} $ 
where $p_p=p_d/2$.
From the weighted average of the experimental measurements of $d(p)$ yields
at 40(20), 30(15), 20(10) and 15(7) GeV/c in the forward direction, $\kappa$ can be
determined as:
$ (0.79 \pm 0.05 \pm 0.13) \times 10^{-2}$,
where the systematic error is dominated by the uncertainty in the 
acceptance calculation.
The dependence of the coalescence factor $\kappa$ as a function of the
momentum of the produced deuteron $p_d$ is shown in figure 3.

\begin{figure}[hbtp!]
\begin{center}
\mbox{\includegraphics[width=.28\linewidth]{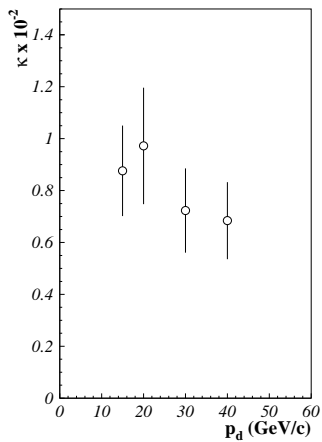}} \\
\end{center}
\caption{  Coalescence factor $\kappa$ as a function of the deuteron 
momentum $p_d$. Errors include statistical and systematics errors.}
\end{figure}




It is interesting to compare this determination with the theoretically 
expected value ($4.6 \times 10^{-2}$ GeV$^2$) from \cite{Braun} (where 
$\kappa$ has no momentum dependence) and the
expectations of reference \cite{Butler} (where a $p_{d}^{-2}$ 
dependence is expected).
Our results are in qualitative agreement with what found in reference 
\cite{Abramov} at lower energies  ($\simeq 1.5 \times 10^{-2}$ GeV$^2$).
According to reference \cite{arsenescu} deuterons are mainly directly
produced in $p+Be$ collisions, while in $Pb+Pb$ collisions  
the dominant production mechanism for deuterons is coalescence.
A possible explanation may be the smaller baryon density and the smaller
source volume available in $p+Be$ collisions compared to $Pb+Pb$ 
collisions.


\end{document}